\documentclass[structabstract]{aa}  
\usepackage{graphicx}
\usepackage{txfonts}
\begin{document}
   \title{The star formation history of the Sculptor Dwarf Irregular Galaxy\thanks{Based on observations made with the NASA/ESA Hubble Space Telescope, obtained from the data archive at the Space Telescope Institute. STScI is operated by the association of Universities for Research in Astronomy, Inc. under the NASA contract NAS 5-26555.}}

   \author{S.~Lianou 
              \inst{1} 
          \and
           A.~A.~Cole
              \inst{2}
                   }
           \institute{National Observatory of Athens, Institute for Astronomy, Astrophysics, Space Applications \& Remote Sensing, Ioannou Metaxa and Vasileos Pavlou, GR-15236 Athens, Greece \\
               \email{s.lianou@astro.noa.gr}
           \and
               School of Mathematics \& Physics, University of Tasmania, Private Bag 37 Hobart, Tasmania 7001, Australia \\ 
               \email{Andrew.Cole@utas.edu.au}
                  }

   \date{Received August 9, 2012; accepted October 26, 2012}

 
  \abstract
   {}
   {We study the resolved stellar populations and derive the star formation history of the SDIG, a gas-rich dwarf galaxy member of the NGC7793 subgroup in the Sculptor group of galaxies.}
   {We construct a color-magnitude diagram using archival observations from the HST/ACS in order to examine the stellar content of SDIG, as well as the spatial distribution of stars selected within different stellar evolutionary phases. We derive the star formation history of SDIG using a maximum-likelihood fit to the color-magnitude diagram.}
   {The color-magnitude diagram shows that SDIG contains stars from 10~Myr to several Gyr old, as revealed from the main sequence, blue loop, luminous asymptotic giant branch, and red giant branch stars. The young stars with ages less than $\sim$250~Myr show a stellar spatial distribution confined to the central regions of SDIG, and additionally the young main sequence stars exhibit an off-center density peak. The intermediate-age and older stars as traced by the red giant branch stars are more spatially extended. SDIG is dominated by intermediate-age stars with an average age of 6.4$^{+1.6}_{-1.4}$~Gyr. The average metallicity inferred from the CMD modelling is [M/H] $\approx$ $-$1.5~dex. SDIG has a star formation history consistent with a constant star formation rate, except for ages younger than $\approx$200~Myr. The lifetime average star formation rate is 1.3$^{+0.4}_{-0.3}$~$\times$10$^{-3}$~M$_{\sun}$~yr$^{-1}$. More recently than 100~Myr, there has been a burst of star formation at a rate $\sim$2--3 times higher than the average star formation rate. The inferred recent star formation rate from CMD modelling, 2.7$(\pm0.5)$~$\times$10$^{-3}$~M$_{\sun}$~yr$^{-1}$, is higher than inferred from the H$\alpha$ flux of the galaxy; we interpret this to mean that the upper end of the initial mass function is not being fully sampled due to the low star formation rate. Additionally, an observed lack of bright blue stars in the CMD could indicate a downturn in star formation rate on 10$^7$-yr timescales. A previous star formation enhancement appears to have occurred between 600-1100~Myr ago, with amplitude similar to the most recent 100~Myr. Older bursts of similar peak star formation rate and duration would not be resolvable with these data. The observed enhancements in star formation suggest that SDIG is able to sustain a complex star formation history without the effect of gravitational interactions with its nearest massive galaxy. Integrating the star formation rate over the entire history of SDIG yields a total stellar mass equal to 1.77$^{+0.71}_{-0.72}$~$\times$10$^{7}$~M$_{\sun}$, and a current V-band stellar mass-to-light ratio equal to 3.2~M$_{\sun}$~/~L$_{\sun}$. }
   {}


   \maketitle
%

\section{Introduction}

   Resolved stellar populations and star formation histories (SFHs) provide a powerful tool to understand galaxy formation and evolution in all scales (e.g., Cignoni \& Tosi \cite{sl_cignoni10}). Detailed SFHs of Local Group (LG) dwarf galaxies have shown that these are complex systems (e.g., Mateo \cite{sl_mateo98}; Tolstoy, Hill \& Tosi \cite{sl_tolstoy09}). The complexity of SFHs observed in LG dwarfs has been explored by means of numerical simulations and the results show that their diversity can be understood by invoking the action of ram pressure stripping, tidal stirring, and/or cosmic ultraviolet (UV) background radiation upon gas-rich dwarf galaxies (e.g., Mayer \cite{sl_mayer10}; Kazantzidis et al.~\cite{sl_kazantzidis11}; Nichols \& Bland-Hawthorn \cite{sl_nichols11}; and references therein). From an observational point of view, the comparison of SFHs between dwarf galaxies in the LG and outside the LG shows that the former are representative of their kind (Weisz et al.~\cite{sl_weisz11a}), so that a deeper study of individual SFHs is valuable to uncover and understand the intrinsic and environmental differences affecting their evolution (e.g., Weisz et al.~\cite{sl_weisz11b}). 

   One such interesting galaxy to focus on is the Sculptor Dwarf Irregular Galaxy (SDIG). First discovered by Laustsen et al.~(\cite{sl_laustsen77}), SDIG is a low-luminosity (Heisler et al.~\cite{sl_heisler97}), gas-rich dwarf irregular (dIrr) galaxy (Cesarsky, Falgarone \& Lequeux \cite{sl_cesarsky77}; Lequeux \& West \cite{sl_lequeux81};  C\^ot\'e et al.~\cite{sl_cote97}). Atomic hydrogen observations show that SDIG has a smooth distribution without the presence of structures such as shells, holes or bubbles (C\^ot\'e et al.~\cite{sl_cote00}). SDIG contains no $\ion {H}{ii}$ regions and this is suggestive of a quiescent phase with very low current star formation rate (SFR; Miller \cite{sl_miller96}; Heisler et al.~\cite{sl_heisler97}; Skillman, C\^ot\'e \& Miller \cite{sl_skillman03}). Bouchard et al.~(\cite{sl_bouchard09}) detect two H$\alpha$ point sources and estimate a very low current SFR (3.8$\times$10$^{-5}$~M$_{\sun}$ yr$^{-1}$); however, Lee et al.~(\cite{sl_lee11}) detect far-ultraviolet (FUV) and near-ultraviolet (NUV) emission that appears to have a clumpy morphology and suggest main-sequence B stars as the source of this emission. 

  With a distance of 3.2~Mpc and based on its radial velocity, SDIG is a member of a loose galaxy triplet (Karachentsev et al.~\cite{sl_karachentsev06}), of which the other two galaxy members are NGC\,7793 at a distance of 3.61$\pm$0.53~Mpc (Vlajic, Bland-Hawthorn \& Freeman \cite{sl_vlajic11}) and UGCA442 at a distance of 3.8$^{+0.8} _{-0.3}$~Mpc (Mould \cite{sl_mould05}). This triplet, the NGC\,7793 subgroup (Karachentsev et al.~\cite{sl_karachentsev03}) lies at the far side of the Sculptor group. SDIG therefore lies near the limit of the ACS Nearby Galaxy Survey Treasury galaxy sample (ANGST; Dalcanton et al.~\cite{sl_dalcanton09}), so a study of the SFH of this subgroup represents an opportunity to learn more about the Sculptor group, a low-density and extended filamentary structure with several embedded subgroups (e.g., Jerjen, Freeman \& Binggeli \cite{sl_jerjen98}). The filamentary structure of the Sculptor group extends from 1.5~Mpc to 4~Mpc, and points to an unvirialised state, with galaxies still falling in (Jerjen, Freeman \& Binggeli \cite{sl_jerjen98,sl_jerjen00}; Karachentsev \cite{sl_karachentsev05}). The early evolutionary state and the low density of the Sculptor group provide an ideal environment in which its galaxy members may be considered as still evolving in isolation.

   In the present work, we examine the resolved stellar populations and model the color-magnitude diagram (CMD) of SDIG using archival Hubble Space Telescope (HST) observations. The structure of this work is as follows. In \S2 we present the observations and photometry, including photometric error analysis. In \S3 we show the CMD and discuss the stellar content of SDIG, including an examination of the spatial distribution of stellar populations selected in several stellar evolutionary phases. In \S4 we briefly review the CMD modelling technique. In \S5 we show the results for the SFH of SDIG. We summarise and discuss our findings in \S6.

\section{Observations and photometry}

   We use archival observations taken with the Advanced Camera for Surveys (SNAP\,9771; Karachentsev et al.~\cite{sl_karachentsev06}) with a total exposure time of 900~sec in the F814W-band, and 1200~sec in the F606W-band. We retrieve the pre-reduced images through the ST-ECF Hubble Science Archive and perform stellar point source photometry using the ACS module of DOLPHOT, a modified version of the photometry package HSTphot (Dolphin \cite{sl_dolphin00}). We apply point source photometry simultaneously on all the individual, calibrated and flat-fielded “FLT” images, and we use as a reference image the drizzled image in the F814W filter. We proceed with the photometry reduction steps as described in the ACS module of DOLPHOT. DOLPHOT provides magnitudes in both the ACS\,/\,WFC and the Landolt UBVRI photometric systems and we choose to use the ACS\,/\,WFC filter system for our study. The final photometric catalogue includes 11363~stars selected to have ${\rm S/N} >$~5, sharpness $|$sharp$_{\rm F606W}+$sharp$_{\rm F814W}|<$1, and crowding $($crowd$_{\rm F606W}+$crowd$_{\rm F814W})<$1 (e.g., Williams et al.~\cite{sl_williams09}). 

   We conduct artificial star tests in order to quantify the photometric errors and completeness of our data, using the utilities provided within DOLPHOT. To that end, we run artificial star tests with 10$^5$~stars per ACS\,/\,WFC chip, with range in F606W from 21~mag to 30~mag, and in (F606W-F814W) from $-$1.5~mag to 3.5~mag. In order to better quantify the photometric errors in the faintest magnitudes, which we need for the CMD modelling, we run complementary artificial star tests with an additional number of 2$\times$10$^5$ stars per ACS\,/\,WFC chip and covering magnitudes fainter than 25~mag. The high number of stars inserted during the artificial star tests is not an issue for self-induced crowding, since each star is inserted and photometered one at a time (Dolphin \cite{sl_dolphin00}; see also Perina et al.~\cite{sl_perina09}). 

 \begin{figure}
  \centering
      \includegraphics[width=8cm,clip]{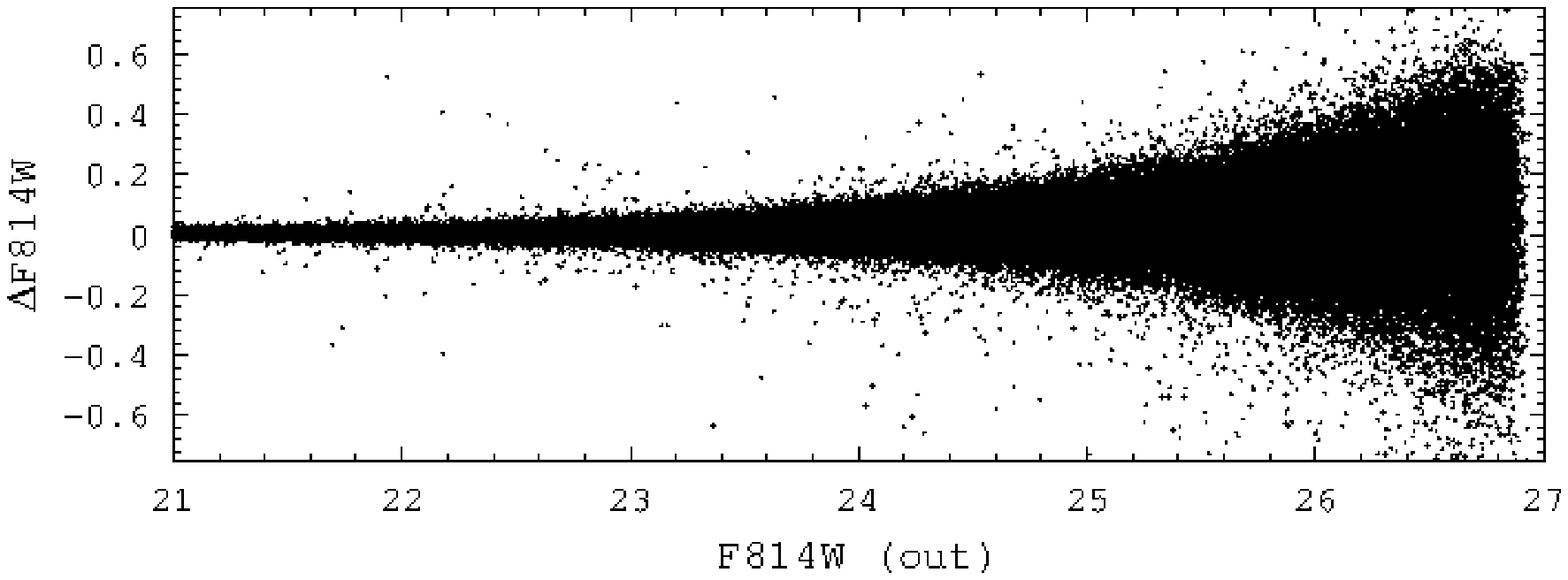}
      \includegraphics[width=8cm,clip]{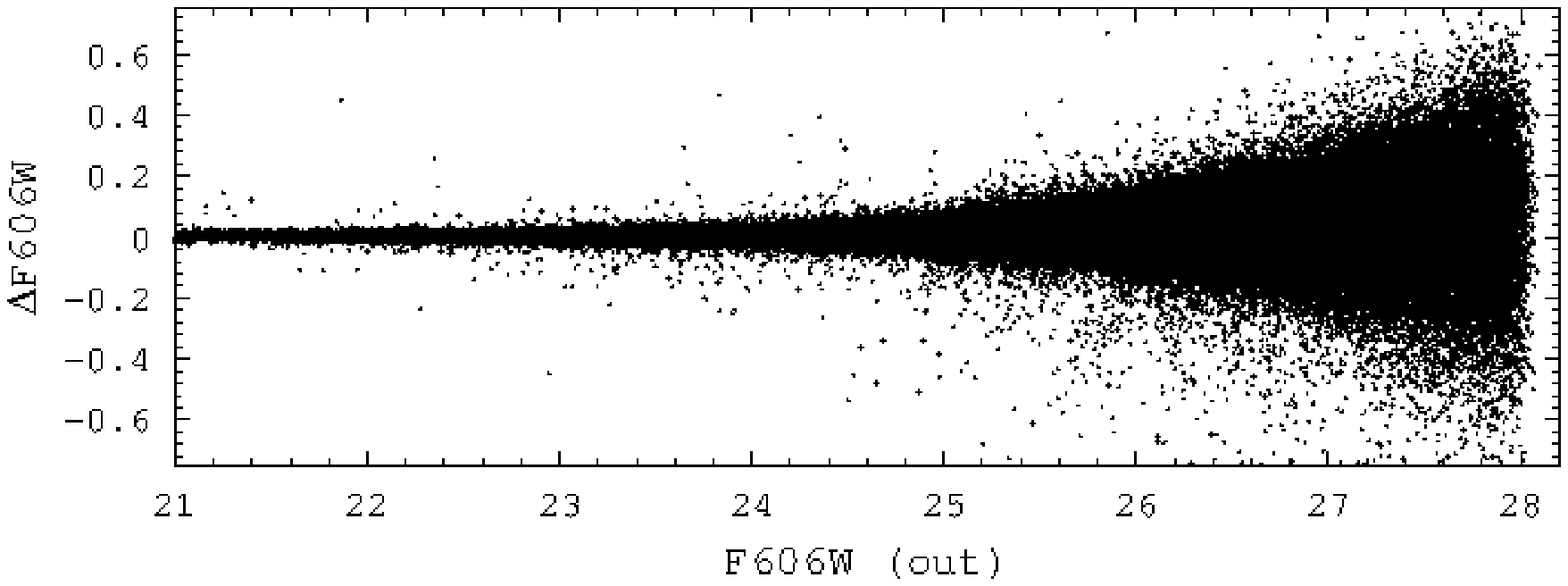}
  \caption{Photometric errors derived from artificial star tests and estimated as the difference between output and input magnitudes, as a function of the output magnitude in the F814W filter (upper panel) and in the F606W filter (lower panel).}
  \label{sl_figure1}%
 \end{figure}
%
  The photometric errors derived using the artificial star tests are shown in Fig.~\ref{sl_figure1}. The mean photometric error and standard deviation is 0.03$\pm$0.1~mag in F814W, and 0.02$\pm$0.09~mag in F606W. 
%
 \begin{figure}
  \centering
      \includegraphics[width=8cm,clip]{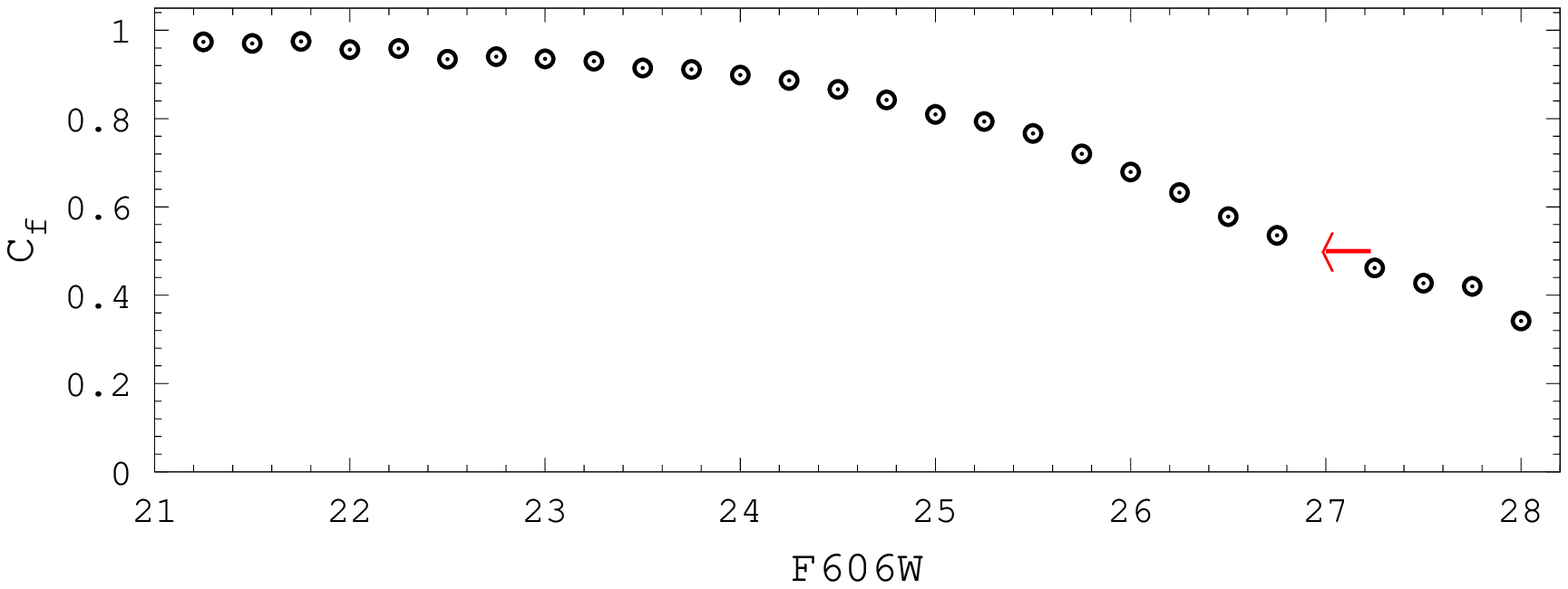}
      \includegraphics[width=8cm,clip]{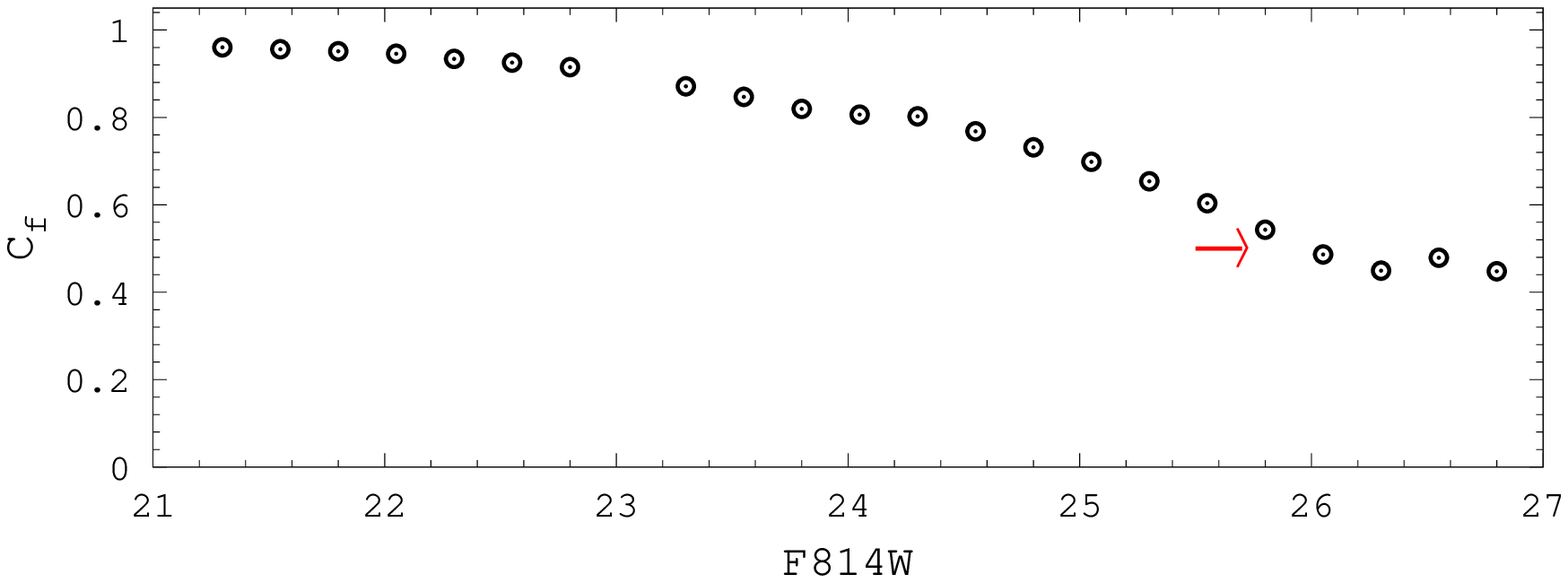}
  \caption{Completeness factors as a function of the F606W-band magnitude, upper panel, and F814W-band magnitude, lower panel. The red arrow in each panel indicates the 50\% completeness factor.}
  \label{sl_figure2}%
 \end{figure}
%
In Fig.~\ref{sl_figure2} we show the completeness factors as a function of magnitude. The 50\% completeness factor occurs at the magnitude F606W~$=\sim$26.9~mag, and F814W~$=\sim$25.9~mag. 

\begin{table}[t]
      \caption[]{Properties of SDIG.}
      \label{table1} 
      \begin{tabular}{lc}
\hline\hline
  Quantity                            & Value                   \\
\hline
  Type                                & dIrr                     \\
  RA~(J2000.0)                        & $00^h08^m13.36^s$         \\
  Dec~(J2000.0)                       & $-34^\circ 34' 42.0''$    \\
  $\rm E(B-V)$~(mag)                  & 0.012                   \\
  $A_{\rm F606W}, A_{\rm F814W}$~(mag)   & 0.034, 0.022            \\
  $F814W_{\rm TRGB}$~(mag)             & 23.44$\pm$0.05          \\
  $(m-M)_0$~(mag)                     & 27.51$\pm$0.06          \\
  Distance~(Mpc)                      &3.2$\pm$0.1              \\
  M$_{\rm V}$~(mag)                   &$-11.87$                 \\
  M$_{\rm {\ion {H}{i}}}$~(M$_{\sun}$)   &2.5$\times$10$^7$        \\
  x$_{\rm C}$~(arcsec)                & 105.1$\pm$1.7           \\
  y$_{\rm C}$~(arcsec)                & 114.5$\pm$1.3           \\
  $e$                                & $0.05$                  \\
  PA~(degrees)                       & 0                       \\
  $r_{\rm eff}$~(arcsec)              & $29.4\pm1.6$            \\
\hline
\end{tabular} 
\end{table}
 Table~\ref{table1} lists the basic properties for SDIG. Each row shows: (1) the galaxy type; (2) and (3) equatorial coordinates of the field center (J2000.0); (4) the reddening adopted from Schlegel et al.~(\cite{sl_schlegel98}); (5) the galactic foreground extinction in the F606W and F814W filters, computed using the extinction ratios for a G2 star listed in Sirianni et al.~(\cite{sl_sirianni05}; their Table~14); (6) the magnitude of the tip of the red giant branch (TRGB) we derive from the data; (7) the distance modulus we derive from the data; (8) the derived distance in Mpc; (9) the absolute V-band magnitude adopted from Georgiev et al.~(\cite{sl_georgiev09}), who use the distance of Karachentsev et al.~\cite{sl_karachentsev06}; (10) the {\ion {H}{i}} mass adopted from Koribalski et al.~(\cite{sl_koribalski04}); (11) and (12) the magnitude-weighted coordinates of the center, x$_{\rm C}$ and y$_{\rm C}$, respectively, which we compute from the data (see $\S$3); (13) and (14) the ellipticity and position angle adopted from Kirby et al.~(\cite{sl_kirby08}); (15) the effective radius adopted from Kirby et al.~(\cite{sl_kirby08}).

\section{Color-magnitude diagram and stellar content}

 \begin{figure}
  \centering
      \includegraphics[width=8cm,clip]{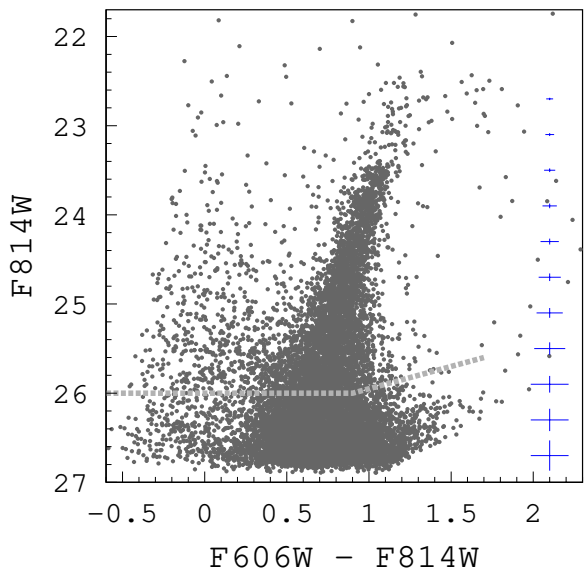}
  \caption{Color-magnitude diagram of SDIG. The gray dashed line shows the 50\% completeness factors. The error-bars correspond to the photometric errors derived using artificial star tests.}
  \label{sl_figure3}%
 \end{figure}
%
   The CMD of SDIG shown in Fig.~\ref{sl_figure3} reveals the presence of several stellar evolutionary features. The most prominent feature is the red giant branch (RGB). The RGB indicates that stars with ages from $\sim$1.5~Gyr and older are present (e.g, Salaris, Cassisi \& Weiss \cite{sl_salaris02}). Intermediate-age stars, with ages between 1~Gyr and 10~Gyr, are revealed in the presence of luminous  asymptotic giant branch (AGB) stars, while the young main-sequence (MS) and blue-loop (BL) stars indicate the presence of stars with ages younger than 1~Gyr. Therefore, SDIG contains stars spanning a wide range in age, from $\sim$Myr to older ages. 

   In the following analyses and unless otherwise noted, we use the TRGB magnitude and distance modulus we measure in the ACS\,/\,WFC filter system. We derive the distance modulus for SDIG using the absolute F814W-band magnitude of the TRGB computed through the calibration of Rizzi et al.~(\cite{sl_rizzi07}): $M_{\rm F814W} = -4.06 + 0.20\times [({\rm F606W-F814W}) - 1.23]$, and the F814W-band magnitude of the TRGB computed through the Sobel-filtering technique on the F814W-band luminosity function (e.g., Lee, Freedman \& Madore \cite{sl_lee93}; Sakai et al.~\cite{sl_sakai96}). We find a TRGB F814W-band magnitude of 23.44$\pm$0.05~mag, while measuring a dereddened color at the TRGB magnitude level of 1.05$\pm$0.03~mag, we derive ${\rm M}_{\rm F814W}=-$4.09$\pm$0.02~mag. The quoted uncertainties take into account the photometric errors from the artificial star tests and the width of the Sobel filter response. These values along with the extinction listed in Table~\ref{table1} give a distance modulus of 27.51$\pm$0.06~mag, which translates to a distance of 3.2$\pm$0.1~Mpc. The quoted uncertainties take into account the width of the Sobel filter response for the TRGB magnitude detection, uncertainties related to the TRGB calibration, and the photometric errors. Our derived distance modulus and distance are consistent with the values of 27.53~mag and 3.21~Mpc, respectively, found in Karachentsev et al.~(\cite{sl_karachentsev06}), who use the same method on the transformed V and I filters and assign an uncertainty to their distance of 8\%. We list the derived F814W-band magnitude of the TRGB, distance modulus, and distance in Table~\ref{table1}.

 \begin{figure*}
  \centering
     \includegraphics[width=6cm,clip]{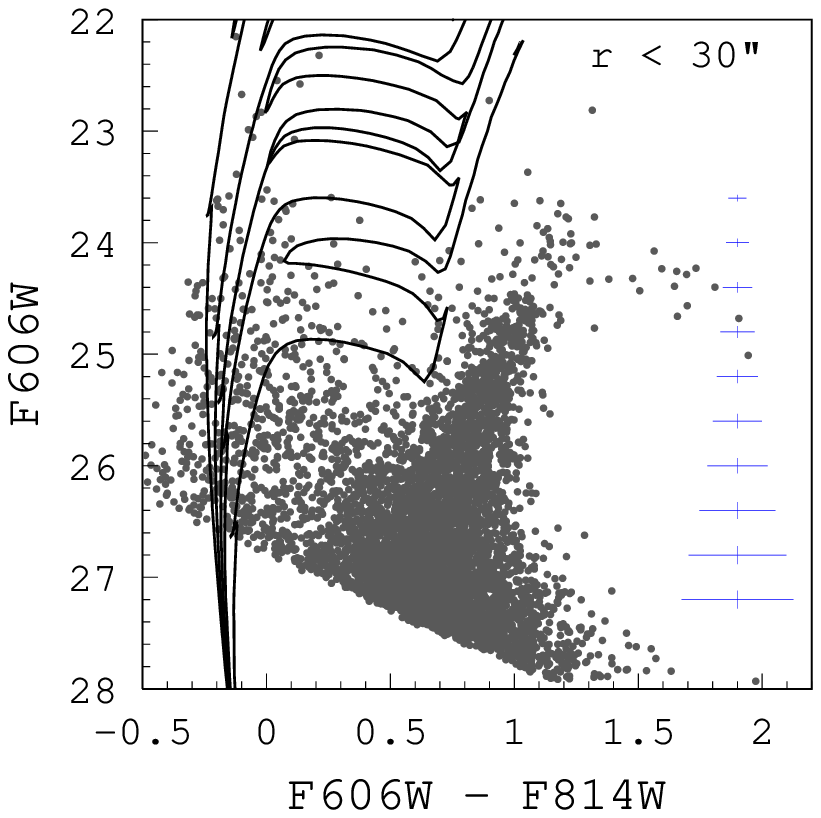}
     \includegraphics[width=6cm,clip]{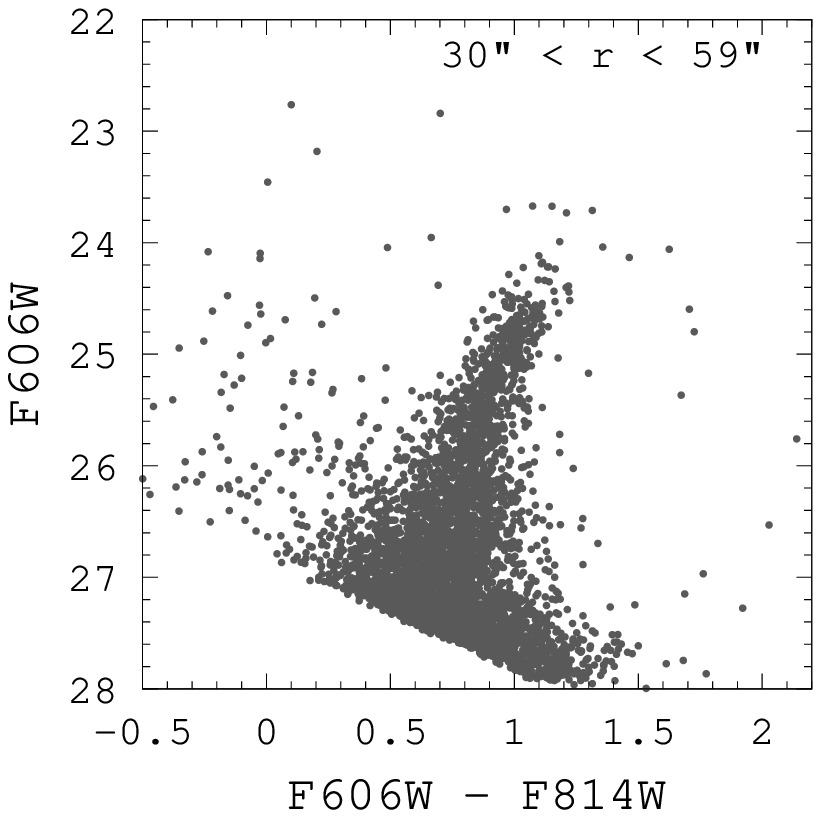}
     \includegraphics[width=6cm,clip]{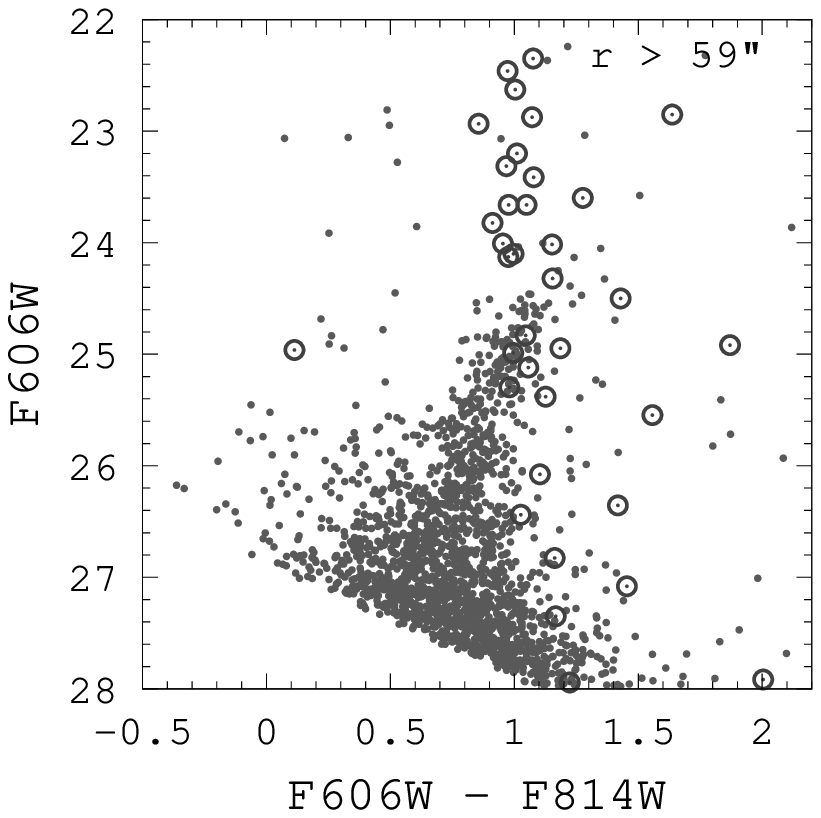}
  \caption{Color-magnitude diagrams within three elliptical radius annuli. The first annulus corresponds to an elliptical radius equal to 1~$r_{\rm eff}$, the second between 1 and 2~$r_{\rm eff}$, and the third beyond 2~$r_{\rm eff}$, shown from left to right, respectively. In the left panel, we overplot Padova isochrones with ages of 20~Myr, 40~Myr, 60~Myr, 80~Myr, and 150~Myr, and for all isochrones a Z metallicity of 0.001. Also shown in the left panel with the error bars the photometric errors. In the right panel, we show with open circles the galactic foreground contamination.}
  \label{sl_figure4}%
 \end{figure*}
%
  Fig.~\ref{sl_figure4} shows the CMD of stars distributed within three elliptical radius annuli: within 1 effective radius, $r_{\rm eff}$; between 1~$r_{\rm eff}$ and 2~$r_{\rm eff}$; and beyond 2~$r_{\rm eff}$. The elliptical radius is defined as: $r = [{\rm x}^{2}+{\rm y}^{2}/(1 - e)^{2}]^{1/2}$, where x and y are the distances along the major and minor axis, respectively, and $e$ is the ellipticity. We adopt the ellipticity, position angle, and effective radius from Kirby et al.~(\cite{sl_kirby08}), listed in Table~\ref{table1}. Also listed in Table~\ref{table1} are the x$_{\rm C}$ and y$_{\rm C}$ coordinates of the magnitude-weighted center of SDIG, which we compute from our data using the equation: ${\rm x_{C}}=\Sigma({\rm mag}_{i}\times {\rm x}_{i})/\Sigma {\rm mag}_{i}$, where ${\rm mag}_{i}$ corresponds to the F814W-band magnitude of each star, and ${\rm x}_{i}$ to their x (or y for the  y$_{\rm C}$) coordinates, in pixels. We use stars with magnitudes brighter than the magnitude with 50\% completeness factors. In the right panel of Fig.~\ref{sl_figure4} we show with the open circles the Galactic foreground contamination, which was estimated using the TRILEGAL code (Girardi et al.~\cite{sl_girardi05}; Vanhollebeke, Groenewegen \& Girardi \cite{sl_vanhollebeke09}). We expect a total of 58 foreground stars in the same color and magnitude location of the SDIG stars, which translates to less than 0.5\% of the total number of SDIG stars. In the left panel of Fig.~\ref{sl_figure4} we overplot Padova isochrones (Marigo et al.~\cite{sl_marigo08}; Girardi et al.~\cite{sl_girardi10}), with ages between 20~Myr and 150~Myr. Fig.~\ref{sl_figure4} indicates that the bulk of the young MS stars are confined within the central parts of SDIG. 

  We examine the cumulative distribution functions of the MS, BL, luminous AGB, and RGB stars, versus the elliptical radius, r, defined in the preceding paragraph. We select those MS stars with their color ranging from $-$0.5~mag to $-$0.1~mag and their F814W-band magnitude ranging from 23.8~mag to 25.25~mag; the BL stars are selected with their color ranging from $-$0.05~mag to 0.35~mag and with their F814W-band magnitude ranging from 21~mag to 25~mag. The RGB stars are selected between the magnitude of the TRGB and 1~mag below, while the color selection ranges such as to follow the slope of the RGB. The luminous AGB stars are selected to be brighter by 0.15~mag than the TRGB (Armandroff et al.~\cite{sl_armandroff93}) and to lie within 1~mag above (${\rm I_{TRGB}}-0.15$)~mag, and within the color range of 1~$<({\rm V-I})_0<$~3.5~(mag). We count a number of 80 luminous AGB stars, which translates into a fraction, i.e. number of luminous AGB stars versus the number of RGB stars within 1~mag below the TRGB magnitude,  N(AGB)~/~N(RGB)=13\%, comparable to that of Sculptor group dwarf galaxies (Lianou et al.~\cite{sl_lianou12}).
 \begin{figure}
  \centering
       \includegraphics[width=6cm,clip]{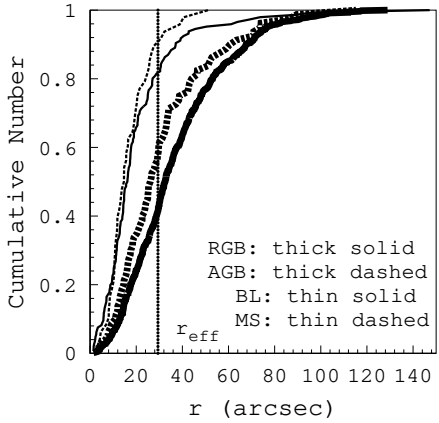}
       \caption{Cumulative distribution functions for the selected stellar populations. The vertical line shows the effective radius, r$_{eff}$.} 
  \label{sl_figure5}%
 \end{figure}
%
The cumulative distribution functions in Fig.~\ref{sl_figure5} show that the MS and BL stars are confined to the central parts of SDIG, while the luminous AGB and the RGB stars are more spatially extended. Performing a two-sided Kolmogorov-Smirnov (K-S) test between each of the cumulative distributions for the MS, BL, and luminous AGB stars with the cumulative distribution of the RGB stars, we derive results consistent with spatially separated populations at the 99\% confidence level in all cases.

 \begin{figure}
  \centering
       \includegraphics[width=4.4cm,clip]{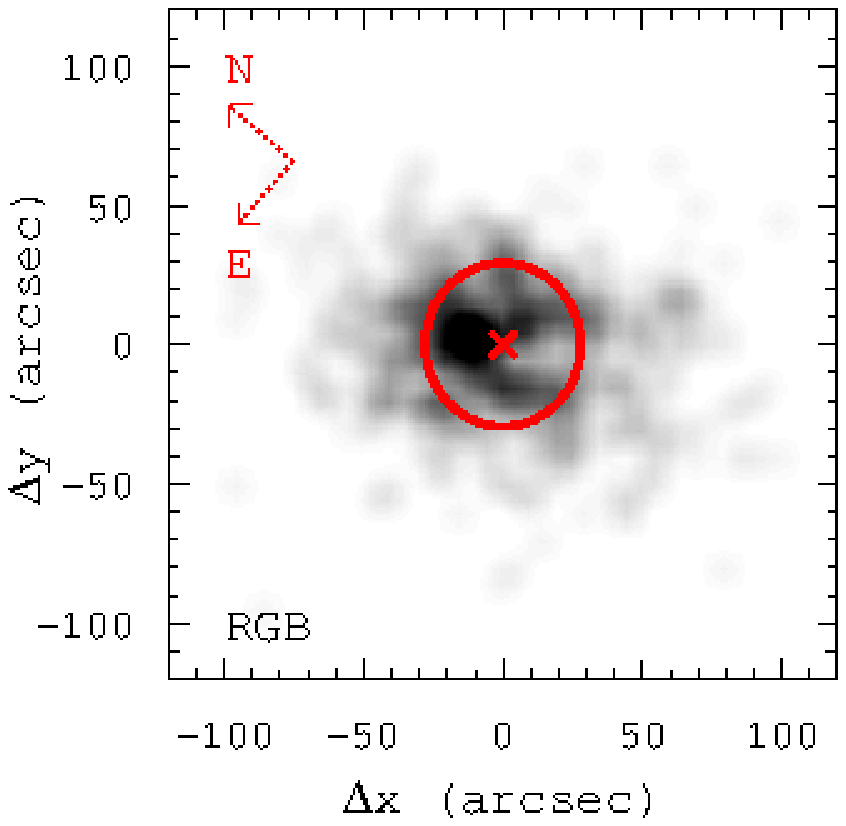}
       \includegraphics[width=4.4cm,clip]{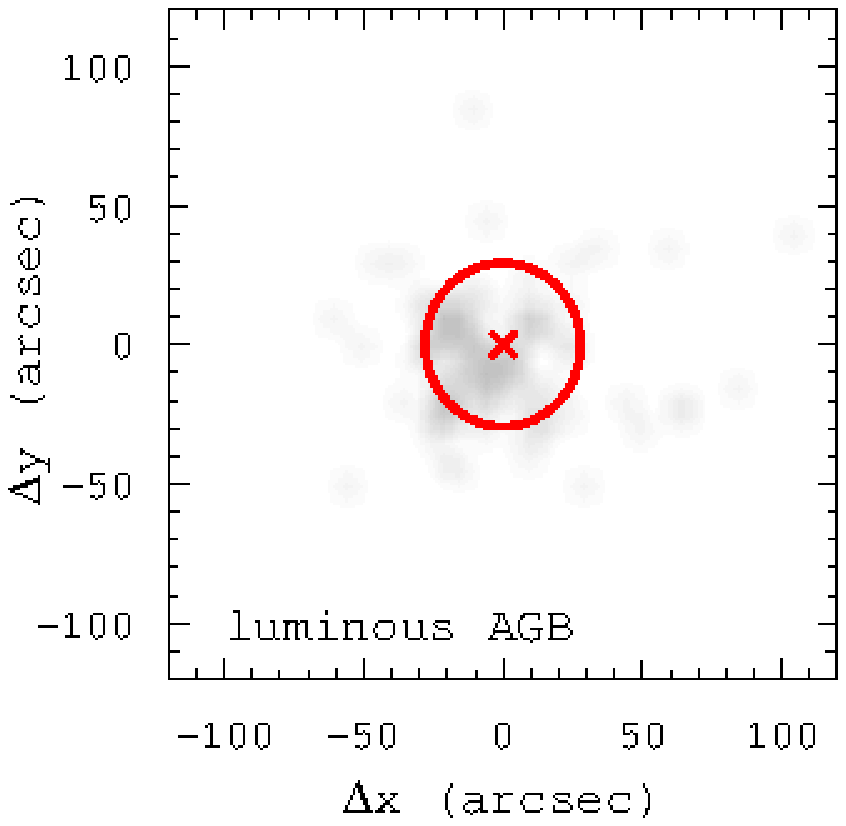}
       \includegraphics[width=4.4cm,clip]{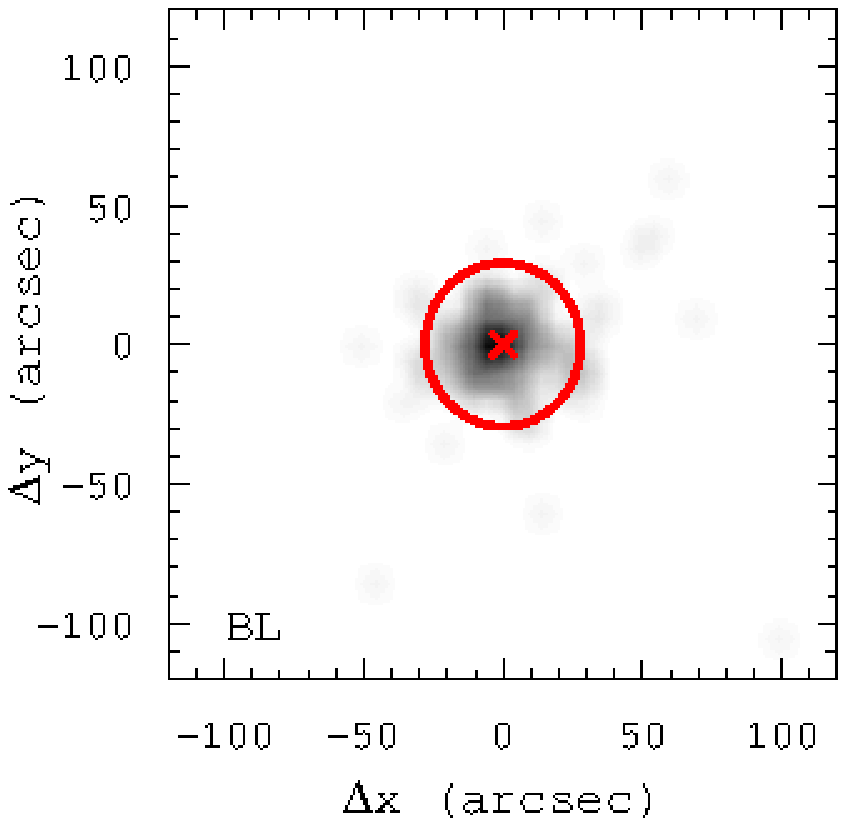}
       \includegraphics[width=4.4cm,clip]{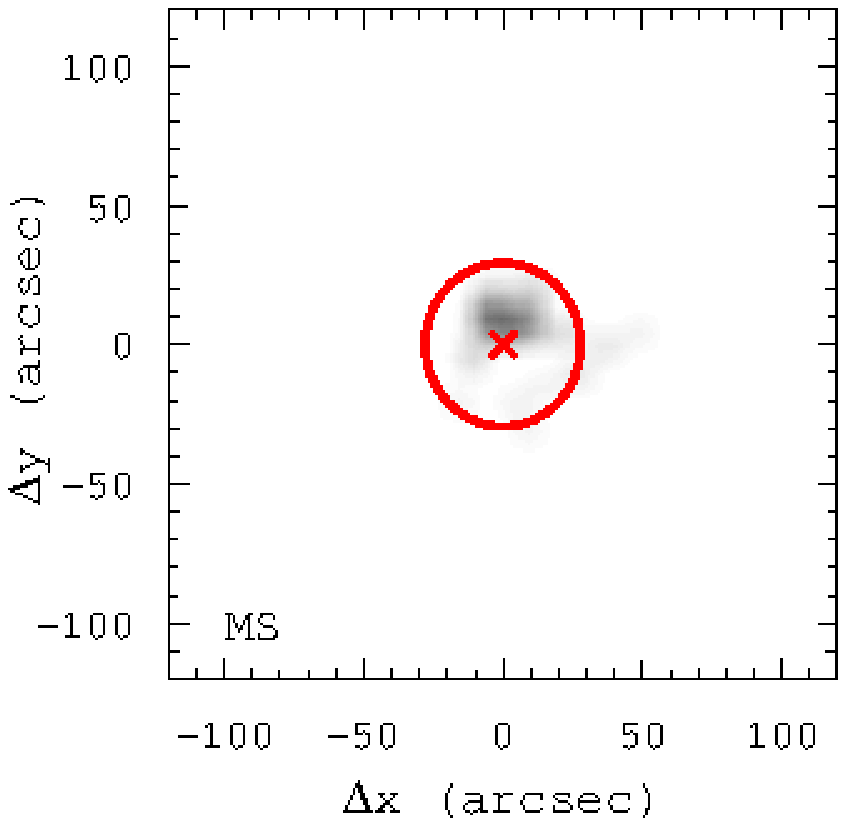}
  \caption{Gaussian-smoothed stellar density maps for stars selected in the MS, BL, luminous AGB, and RGB phases, expressed in terms of angular offsets from the dwarf's center. The density maps are color coded such that black corresponds to a density of 0.16 stars per sq.~arcsec and the density declines toward lighter grey. The ellipse has a major axis equal to 1~$r_{\rm eff}$, while the crosses emphasise the location of the center of the dwarf. The red arrows in the panel of the RGB population indicates the directions of North and East, and is the same for all panels.}
  \label{sl_figure6}%
 \end{figure}
%
  Comparing the spatial distribution of the luminous AGB stars with that of the RGB stars, in the stellar density maps shown in Fig.~\ref{sl_figure6}, the former are more confined in the inner parts of SDIG. The selected BL stars are confined to the central parts of the dwarf when we compare their spatial distribution with the one of the luminous AGB and RGB stars. The stellar density map of the young MS stars shows a population concentrated towards the central parts of SDIG, while in the same time its spatial location is not coincident with the very center of SDIG, having an offset of 8.2~arcsecs, or 130~pc at a distance of 3.2~Mpc. We note that a similar characteristic of off-center local sites of young star formation characterise the majority of the transition-type dwarfs in the Sculptor group (Jerjen \& Rejkuba \cite{sl_jerjen01}; Lianou et al.~\cite{sl_lianou12}), while the same is also observed for the young stellar populations of other LG dwarf galaxies (e.g, Dohm-Palmer et al.~\cite{sl_dohm-palmer97}; Gallagher et al.~\cite{sl_gallagher98}; Mateo \cite{sl_mateo98}; Cole et al.~\cite{sl_cole99}; McConnachie et al.~\cite{sl_mcconnachie06}).

\section{Color-magnitude diagram modelling}

  The basic idea using the modelling of the CMD to uncover the evolutionary history of a galaxy is that its composite stellar populations can be regarded as a linear combination of simple stellar populations of different ages and metallicities. Thus, the SFH can be obtained by comparing the observed features in the CMD with those in a model CMD. There are several features in a CMD that can be used as age and metallicity indicators (e.g., Gallart, Zoccali \& Aparicio \cite{sl_gallart05}). Synthetic populations are constructed using a set of stellar evolution models, adopting an initial mass function (IMF) and applying a star formation and a chemical enrichment law (e.g., Gallart, Zoccali \& Aparicio \cite{sl_gallart05}; Cignoni \& Tosi \cite{sl_cignoni10}). In practice, the density of stars within a defined color and magnitude grid along the observed CMD is compared with the density of the synthetic stars drawn from the model CMD (for a review see Tolstoy, Hill \& Tosi \cite{sl_tolstoy09}; Aparicio \cite{sl_aparicio02}), and the best model is adopted as the SFH.  

  The approach taken here was developed by Cole to study the stellar populations of dIrr galaxies in the LG (see, e.g., Skillman et al.~\cite{sl_skillman03b}; Cole et al.~\cite{sl_cole07}; Cignoni et al.~\cite{sl_cignoni12}). As with any such approach, the power of the method is greatly reduced by lack of photometric depth (e.g., Weisz et al.~\cite{sl_weisz11b}). Detailed inferences about the SFH will be restricted to age ranges for which the age-sensitive MS, BL, and to a lesser extent AGB stars are visible. It is still possible to place constraints on long-term average SFRs and mean metallicities based on inferences from the RGB and the regions of the CMD that are badly affected by incompleteness. However, these inferences are by necessity more model-dependent and of lower precision, and the resulting uncertainties are dominated by systematic effects.

  The SFH-fitting starts with theoretical isochrones interpolated to a fine grid of age and metallicity in order to create a synthetic CMD with no gaps. The most recent models from the Padova models are used. The synthetic CMDs are binned in age and metallicity (Z) to avoid ``overfitting'' noise in the CMDs. The bins are initially spaced by a default value of $\Delta$log(age) = 0.10, but adjacent bins are merged when lack of information in the CMDs leads to instability in the solutions. In this case, ages greater than 1~Gyr are only represented on the RGB and AGB, and the age resolution is decreased to $\Delta$log(age) = 0.30 (factor of two). There are not many stars younger than 30~Myr in the CMD, so the youngest age bin spans the entire range from 4--30~Myr as a way to populate the CMD bins sufficiently to draw SFH inferences.  

  The CMD is divided into a regular grid of color-magnitude cells, and the expectation value of the number of stars in each cell for a SFR of 1 M$_{\odot}$ yr$^{-1}$ is calculated from the isochrones.  Several parameters are assumed to be fixed during the solution: the distance modulus and reddening, IMF, fraction of binaries, and the binary mass ratio distribution function. The adopted IMF is from Chabrier (\cite{sl_chabrier03}), and the binary fraction and mass ratios are parameterized based on Duquennoy \& Mayor (\cite{sl_duquennoy91}) and Mazeh et al.~(\cite{sl_mazeh92}). We take 35\% of stars to be single and the rest to be binary. The binaries are divided into ``wide'' and ``close'' binaries in a 3:1 ratio; the secondaries in the wide systems are drawn from the same IMF as the primaries, but in the close systems the secondary masses are drawn from a flat IMF. The distance and reddening are initially constrained to the values given in Table~1, but are varied if the resulting synthetic CMDs are mismatched to the data.

  No age-metallicity relation is explicitly assumed, but a range of metallicities at each age is allowed, constrained by the color range of the data. Because of the shallow photometric depth of the data, there are few strong constraints on the metallicity, and those that exist are model-dependent. The RGB is affected by age-metallicity degeneracy, but indicates a range of metallicities between 0.0004 $\leq$ Z $\leq$ 0.002. These numbers are consistent with the Padova AGB star tracks, and the color separation between the BL stars and MS. There is no metallicity constraint from $\ion {H}{ii}$ regions, but the inferred values are consistent with the metallicity-luminosity relationship for dIrr galaxies in Lee, Zucker \& Grebel (\cite{sl_lee07}) (see discussion below).

  The IMF-weighted, color-magnitude binned isochrones are convolved with color and magnitude error distributions derived from the artificial star tests in order to create synthetic CMDs with the same properties as the data. Linear combinations of the synthetic CMDs are created by the fitting routine, which uses a maximum likelihood test based on the Poisson distribution of counts in each bin (Cash \cite{sl_cash79}) to find the combination that is most likely to produce the observed CMD. We use a simulated annealing technique to find the best fit while avoiding false, local maxima in likelihood space. The SFH errorbars are calculated by successive perturbation of each age-metallicity bin; because the total number of stars is fixed by the observations, a decrease in SFR in one bin results in an increase in adjacent bins; this gives anticorrelations between adjacent bins. In general, the code is driven by the most populous cells of the CMD to its preferred solution; because most stars are observed on the MS and low-mass stars vastly outnumber those of higher mass, the most significant cells in determining the SFH are frequently below the 50\% completeness threshold, and the solution depends sensitively on the artificial star test results.

  Our data are not deep enough to reach the oldest MS turn-offs, and due to the age-metallicity degeneracy on the RGB, the solution of the CMD modelling for ages higher than $\sim$1.5~Gyr bears larger uncertainties. 
 \begin{figure*}
  \centering
      \includegraphics[width=17cm, clip]{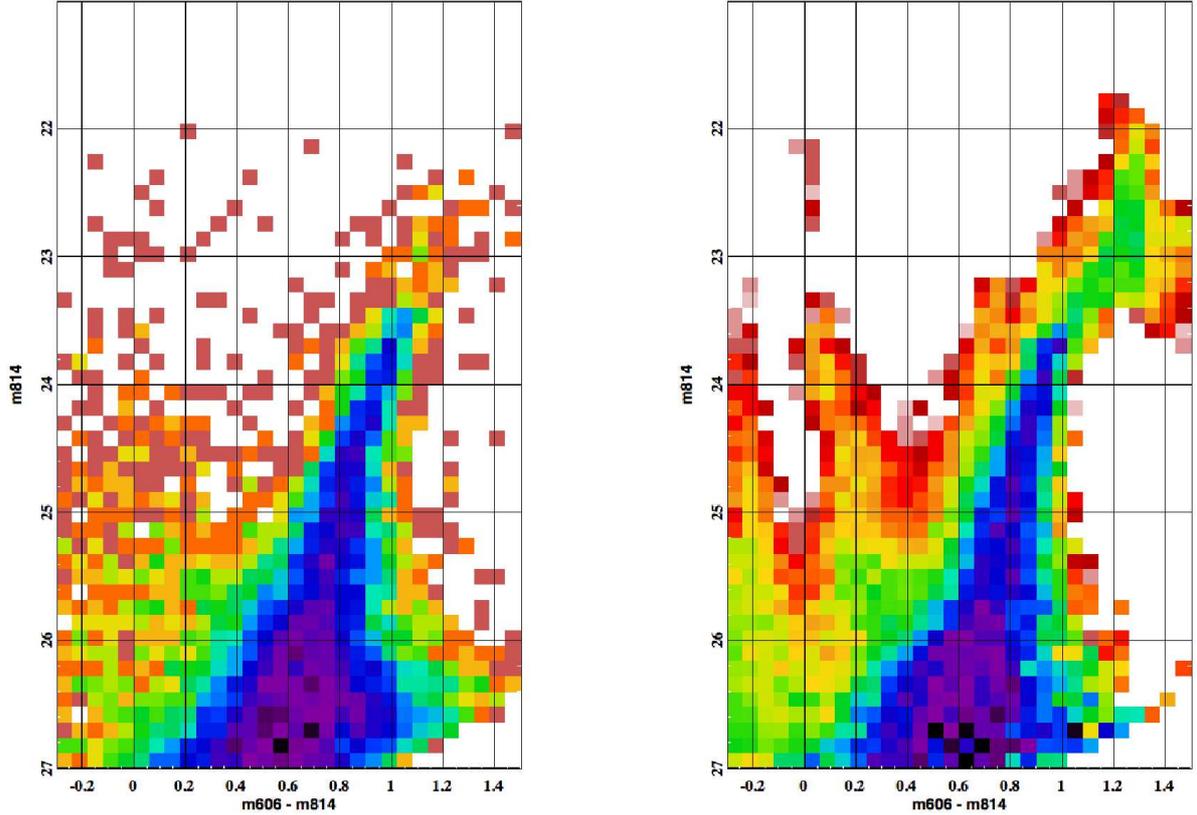}
  \caption{Left panel: logarithmically-scaled, binned CMD used in the maximum-likelihood fit for SFH; right panel: best-fit model. Without differential reddening, there is clear separation between the MS and BL stars, which are mingled together in the data. The AGB stands out as being poorly-fit; the mismatch occurs because of the large number of stars fainter than m814 $\approx$26, which are part of the same stellar population as the AGB stars.}
  \label{sl_figure7}%
 \end{figure*}
%
The observed and modelled CMDs for SDIG are shown in Fig.~\ref{sl_figure7}. From the CMD modelling, we derive a distance modulus of 27.54~mag, while the reddening required is E(B-V)=0.04. The higher reddening, as compared to the value from Schlegel et al.~(\cite{sl_schlegel98}), is consistent with the one that Heisler et al.~(\cite{sl_heisler97}) derive for the very central parts of SDIG.

  The best-fit solution under the restrictions of the fit procedure matches the mean stellar density and colors of the major stellar sequences well, but there are some notable differences between the simulated CMD and the data. The MS and BL in the data are blurred, with no clear color gap between them; however, in the models these features are well-separated. Presumably this is due to differential reddening, with a tendency for young stars to be associated with dust clouds and circumstellar material. 

 \section{Star formation history}

 \begin{figure}
  \centering
      \includegraphics[width=8cm,clip]{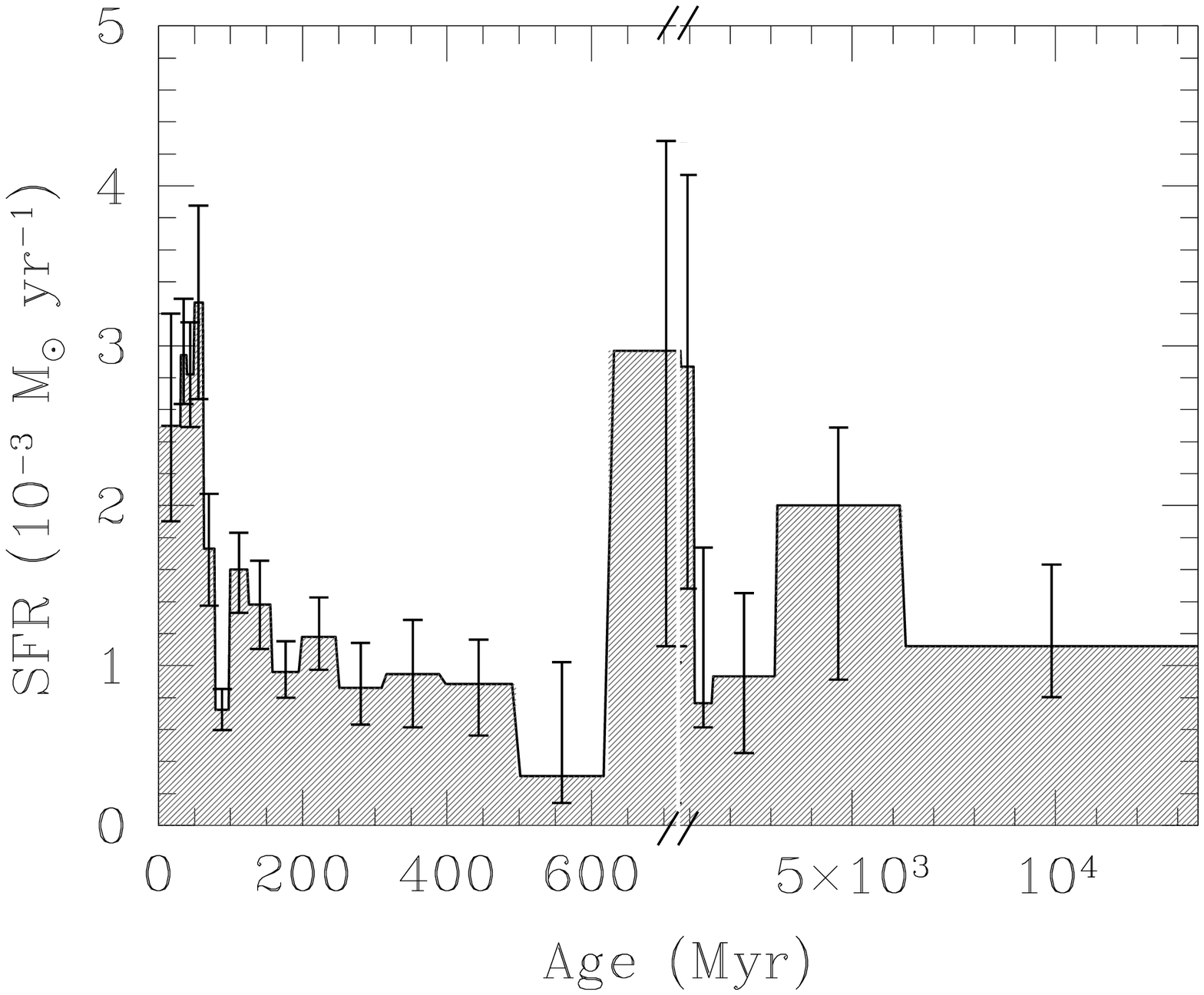}
      \includegraphics[width=8cm,clip]{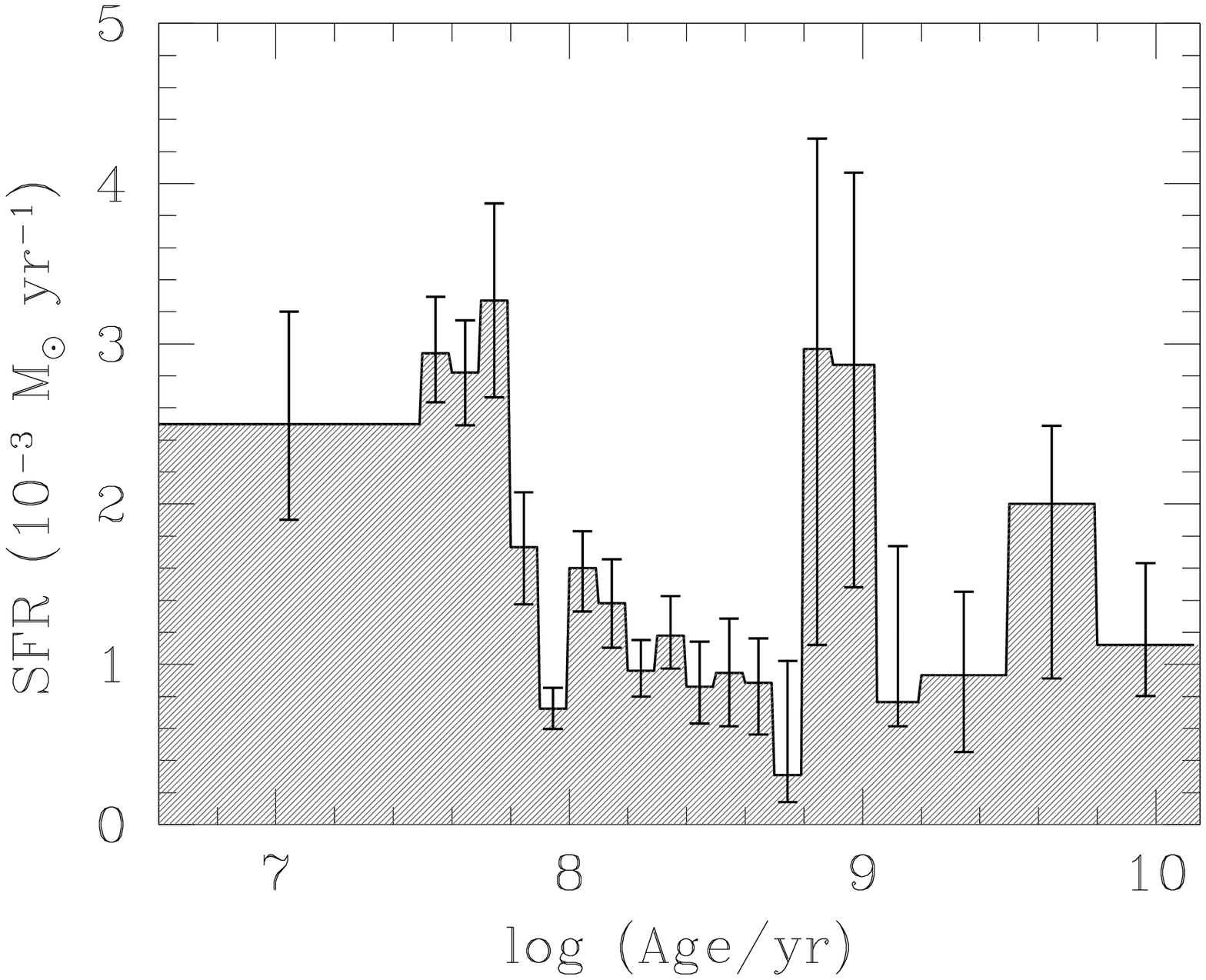}
  \caption{Star formation rate as a function of age for SDIG based on the maximum-likelihood fit, shown in the upper panel with a linear scale on age, and in the lower panel with a logarithmic scale with age. Note in the upper panel the break in age scale at 750~Myr that reflects the decreasing time resolution of the CMD with increasing age.}
  \label{sl_figure8}%
 \end{figure}
%
  We show the best-fit SFH for SDIG in Fig.~\ref{sl_figure8}. Overall, the best-fit SFH is characterised by the following features:

\begin{itemize}

\item An average age for formation of stars of $\langle \tau \rangle$ = 6.4$^{+1.6}_{-1.4}$~Gyr and a total astrated mass (time-integrated SFR) of 17.7$\times$10$^6$M$_{\sun}$.

\item The average metallicity is inferred to be Z $\approx$0.0006 ([M/H] $\approx$ $-$1.5~dex), with some tendency to increase with time, as the BL stars are more consistent with a metallicity closer to Z = 0.001.

\item A long-term SFH that is consistent with a constant SFR over several Gyr. The average SFR is 1.3$^{+0.4}_{-0.3}$~$\times$10$^{-3}$~M$_{\sun}$~$yr^{-1}$. Any bursts or fallow periods are of short enough duration and/or amplitude that they do not strongly perturb the mean SFR averaged over factor of two periods in age. This is broadly consistent with the typical dIrr galaxy found in Weisz et al.~(\cite{sl_weisz11b}); however the HST SDIG observations analysed here were not deep enough to reveal factor of 2--3 variation in SFR over sub-Gyr intervals at intermediate or old ages, and these are not excluded by the CMD analysis.

\item The age resolution improves dramatically for ages less than $\approx$800--1000~Myr, but strong variations over intervals of a few hundred Myr are not observed, with two exceptions:

\begin{enumerate}
\item Over the past 100~Myr the mean SFR increases by a factor of $\approx$2 over its long-term average value, but then declines slightly. Variations of this duration and amplitude would be completely hidden by photometric incompleteness and age-metallicity degeneracy for ages older than $\approx$1~Gyr; this behaviour may be typical of the history of low-luminosity dIrrs (e.g., Cole \cite{sl_cole10}). Because the fraction of evolved stars is low, the derived SFR is susceptible to uncertainties in the modelling of incompleteness and possible variations in the IMF. Stars in this age range would contribute to the SFR as measured from FUV data, but not to the H$\alpha$ luminosity owing to the low absolute SFR with a value of $\sim$2.5$^{+0.7}_{-0.6}$~$\times$10$^{-3}$~M$_{\sun}$~$yr^{-1}$ from 0--10~Myr.

\item A similar-amplitude SFR enhancement is inferred to have taken place from 600--1100~Myr ago. The younger stars in this interval are directly observable in the CMD as BL stars at F606W-band magnitude $\approx$ 26.5--27~mag on the blue side of the RGB. The older stars are confused with the RGB and highly subject to incompleteness, resulting in the large uncertainties in the SFR at this age range. Because of the high incompleteness, the best-fit SFR has high uncertainty (reflected in the model overproduction of AGB stars associated with this age range). Based on this data alone the balance of evidence favours the existence of a $\lesssim$500-Myr long episode of increased SFR $\approx$1~Gyr ago, but is consistent with a flat SFH given the small number of stars and the statistical nature of the problem. As above, this type of event would be undetectable at older ages based on this data set alone. 
\end{enumerate}
\end{itemize}

  Integrating the SFR over the entire history gives a total astrated stellar mass M$_{\star}$=~1.77$^{+0.71}_{-0.72}$~$\times$10$^{7}$~M$_{\sun}$. For ages less than 1~Gyr, we estimate a stellar mass M$_{\star, <1 {\rm Gyr}} =$~1.68$^{+0.69}_{-0.75}$~$\times$10$^{6}$~M$_{\sun}$, which consists a fraction of $\sim$10\% of the total stellar mass. Because the maximum mass of surviving stars decreases with age, only $\approx$85\% of the total astrated mass is still present in the form of stars, i.e., the current stellar mass M$_{\star, {\rm current}} =$~1.50$^{+0.71}_{-0.72}$~$\times$10$^{7}$~M$_{\sun}$. Kirby et al.~(\cite{sl_kirby08}) derive a total stellar mass equal to  1.6$\times$10$^{7}$~M$_{\sun}$, using deep near-infrared observations that trace the old stellar populations and assuming an H-band stellar mass-to-light ratio of 1. The agreement between our current stellar mass and the total stellar mass of Kirby et al.~(\cite{sl_kirby08}) is very good. 

  The current stellar mass is comparable to the {\ion {H}{i}} mass in SDIG (see Table~\ref{table1}; Koribalski et al.~\cite{sl_koribalski04}) with a ratio of M$_{\rm {\ion {H}{i}}}$~/~M$_{\star} =$~0.6. We compute the baryonic gas fraction, M$_{\rm gas}$~/~M$_{\rm baryonic}$, and compare it with the baryonic gas fraction of the ANGST sample galaxies (Weisz et al.~\cite{sl_weisz11b}; their Fig.~9). The baryonic gas fraction of SDIG is estimated equal to M$_{\rm gas}$~/~M$_{\rm baryonic} =$0.7, assuming a gas mass M$_{\rm gas}=$1.4$\times$M$_{\rm {\ion {H}{i}}}$ and a baryonic mass M$_{\rm baryonic}=$M$_{\rm gas} +$M$_{\star}$. These values are listed in Table~\ref{table2}. Placing SDIG in Fig.~9 of Weisz et al.~(\cite{sl_weisz11b}) locates it in the upper left region occupied by dIrrs, with SDIG slightly above ESO321-G014. 

  Based on our current stellar mass estimate and the V-band absolute magnitude listed in Table~\ref{table1}, we compute a V-band stellar mass-to-light ratio equal to M$_{\star}$~/~L$_{\rm V} =$ 3.2~M$_{\sun}$~/~L$_{\sun}$. This is higher than transition-type LG galaxies such as LGS3 and DDO\,216 (Hunter, Elmegreen \& Ludka \cite{sl_hunter10}). Using the B-band magnitude listed in Karachentsev et al.~(\cite{sl_karachentsev04}; column (6) in their Table~1), we compute a B-band stellar mass-to-light ratio equal to M$_{\star}$~/~L$_{\rm B} =$1.4~M$_{\sun}$~/~L$_{\sun}$. The B-band stellar mass-to-light ratio of SDIG is lower than the average B-band stellar mass-to-light ratio (equal to 2.4~M$_{\sun}$~/~L$_{\sun}$) of the sample of star-burst dwarf galaxies listed in McQuinn et al.~(\cite{sl_mcquinn10}), but is higher than the average value of $\sim$1.1~M$_{\sun}$~/~L$_{\sun}$ for dIrrs with $-$11~mag $\leq$ M$_B$ $\leq$ $-$13~mag (Weisz et al.~\cite{sl_weisz11b}). 

  SDIG is a late-type dwarf galaxy and the dominant population has intermediate ages, with an average age of 6.4$^{+1.6}_{-1.4}$~Gyr. Thus, the age-metallicity degeneracy on the RGB makes it difficult to derive individual stellar metallicities. On the other hand, the {\em mean} stellar metallicity of mixed stellar populations remains rather robust against the age-metallicity degeneracy on the RGB (Lianou et al.~\cite{sl_lianou11}). Therefore, we use the average age of the stellar populations in SDIG 
 \begin{figure}
  \centering
     \includegraphics[width=7.5cm,clip]{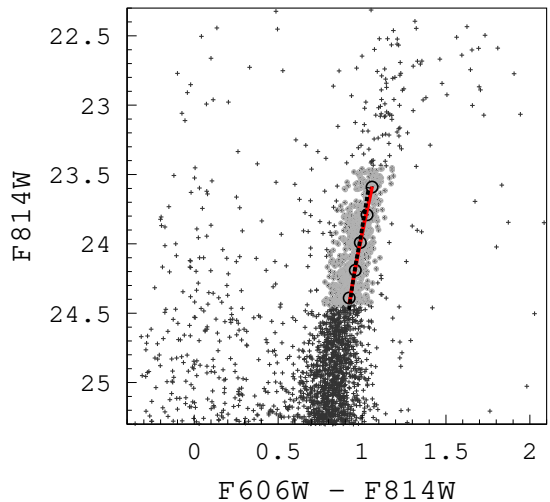}
  \caption{CMD focused on the RGB stars. The grey circles indicate the RGB stars used to compute the fiducial line, with the latter shown with the open black circles. The solid red line shows the Dartmouth 6.4~Gyr, solar-scaled isochrone with metallicity $-$1.6~dex in [M/H] that best matches the RGB fiducial, while the black dotted line shows the Padova 6.4~Gyr isochrone with metallicity $-$1.5~dex in [M/H]. Towards fainter magnitudes, the two isochrones overlap. }
  \label{sl_figure9}%
 \end{figure}
%
in order to constrain its {\em mean} stellar metallicity. To that end, we use the fiducial of RGB stars, with magnitudes ranging from the TRGB to 1~mag fainter and shown in Fig.~\ref{sl_figure9} with the open circles, to which we overplot the best-matched isochrone from the Dartmouth (Dotter et al.~\cite{sl_dotter07,sl_dotter08}) and Padova Stellar Evolution Databases, with an age equal to the average age of the stellar population in SDIG. Here, we use the colour excess and distance modulus we derive from the SFH modelling (see $\S$5). The best-matched isochrones are shown in Fig.~\ref{sl_figure9}, with the red solid line and black dotted line, respectively, for the Dartmouth and Padova isochrones. The metallicity of the best-matched Dartmouth isochrone is $-$1.6~dex, while the metallicity of the best-matched Padova isochrone is $-$1.5~dex. The metallicity of the Padova isochrone that best matches the RGB fiducial is consistent with the average metallicity we derive from the SFH-modelling. The inferred stellar metallicity translates to an [O/H] value of $-$1.13~dex (Mateo \cite{sl_mateo98}), or to a 12+${\rm log}$[O/H] value of 7.78, therefore is consistent with the metallicity-luminosity relation in Lee et al.~(\cite{sl_lee07}). Moreover, the stellar metallicity of SDIG is consistent with the stellar metallicity-luminosity relation in Cloet-Osselaer et al.~(\cite{sl_cloet-osselaer12}). 

\begin{table}[t]
      \caption[]{Derived properties of SDIG.}
      \label{table2} 
      \begin{tabular}{lc}
\hline\hline
 Quantity                           & Value                                                       \\
\hline
$\langle {\rm SFR} \rangle$         & 1.3$^{+0.4}_{-0.3}$~$\times$10$^{-3}$~M$_{\sun}$~$yr^{-1}$     \\
SFR$_{\rm current}$                   & 2.5$^{+0.7}_{-0.6}$~$\times$10$^{-3}$~M$_{\sun}$~$yr^{-1}$     \\
SFR$_{\rm recent}$                    & 2.7$\pm$0.5$\times$10$^{-3}$~M$_{\sun}$~$yr^{-1}$             \\
$\langle {\rm \tau} \rangle$        & 6.4$^{+1.6}_{-1.4}$~Gyr                                        \\ 
M$_{\star}$                          & 1.77$^{+0.71}_{-0.72}$~$\times$10$^{7}$~M$_{\sun}$              \\
M$_{\star, <1 {\rm Gyr}}$              & 1.68$^{+0.69}_{-0.75}$~$\times$10$^{6}$~M$_{\sun}$              \\
M$_{\star, {\rm current}}$             & 1.50$^{+0.71}_{-0.72}$~$\times$10$^{7}$~M$_{\sun}$              \\
M$_{\rm gas}$                         & 3.5$\times$10$^{7}$~M$_{\sun}$                                \\
M$_{\rm baryonic}$                     & 5$\times$10$^{7}$~M$_{\sun}$                                  \\
M$_{\rm gas}$~/~M$_{\rm baryonic}$      & 0.7                                                          \\
M$_{\star}$~/~L$_{\rm V}$              & 3.2~M$_{\sun}$~/~L$_{\sun}$                                    \\
M$_{\star}$~/~L$_{\rm B}$              & 1.4~M$_{\sun}$~/~L$_{\sun}$                                    \\
${\rm [M/H]}_{\rm SFH}$               & $-$1.5~dex                                                    \\
${\rm [M/H]}_{\rm Padova}$             & $-$1.5~dex                                                    \\
${\rm [M/H]}_{\rm Dartmouth}$          & $-$1.6~dex                                                    \\
\hline
\end{tabular} 
\end{table}
The derived properties of SDIG are listed in Table~\ref{table2}. Each row lists the following information: (1) the average SFR; (2) and (3) the current and recent SFRs, respectively (see below); (4) the average age of the stellar populations; (5) the astrated stellar mass; (6) the stellar mass for stars younger than 1~Gyr; (7) the current stellar mass, M$_{\star, {\rm current}}$; (8) the gas mass, computed as 1.4$\times$M$_{\rm {\ion {H}{i}}}$; (9) the baryonic mass, computed as M$_{\rm gas}+$M$_{\star, {\rm current}}$; (10) the baryonic gas fraction, computed as M$_{\rm gas}/$M$_{\rm baryonic}$; (11) the current V-band stellar mass-to-light ratio; (12) the current B-band stellar mass-to-light ratio; (13) the average metallicity based on the SFH results; (14) the stellar metallicity based on the RGB fiducial and Padova isochrones; (15) the stellar metallicity based on the RGB fiducial and Dartmouth isochrones. 

We can compare the SFR in our youngest age bin to current SFR indicators like H$\alpha$ and FUV emission.
The current SFR is estimated from our data assuming it is equal to the average SFR in our first age bin (4--30~Myr), 2.5$^{+0.7}_{-0.6}$~$\times$10$^{-3}$~M$_{\sun}$~$yr^{-1}$. H$\alpha$ emission is sensitive to SFR over the past 10~Myr but can be biased to lower values than the true current SFR when the SFR is less than $\approx$3$\times$10$^{-3}$~M$_{\odot}$~yr$^{-1}$ (e.g., Lee et al.~\cite{sl_lee09}) because it relies on ionising photons from O stars, which may not be fully sampled by the IMF in cases of low SFR.  We detect of order a dozen stars bright enough and blue enough to be candidate OV stars, so stochastic effects are expected to be important. Despite this caveat, the current SFR we derive from the CMD modelling is consistent with the upper limit of 2$\times$10$^{-3}$~M$_{\sun}$~$yr^{-1}$ that Heisler et al.~(\cite{sl_heisler97}) derive based on their 3~$\sigma$ detection of the H$\alpha$ surface brightness across the continuum-subtracted image of SDIG.  However, Bouchard et al.~(\cite{sl_bouchard09}) derive a lower current SFR of 3.8$\times$10$^{-5}$~M$_{\sun}$~yr$^{-1}$ based on the H$\alpha$ flux in two detected knots. Their SFR is just 3\% of the value we find. This could be taken as evidence for a quiescent period of low SFR, or for random variance due to the low numbers of massive stars formed. 

We can estimate the number of O-type stars expected to be produced by the recent SFR using the stellar population synthesis code Starburst99 (Leitherer et al.~\cite{sl_leitherer99}). As a rough estimate, we assume continuous star formation at a constant rate equal to the recent SFR (see Table~\ref{table2}), a Salpeter IMF, and solar metallicity. The results provide us with the total number of ionising photons emitted during the most recent time step of 2~Myr, and over a longer period of 100~Myr. Using the ionising flux Q$_0$ for an O5V star (Q$_0$ = 10$^{49.22}$ s$^{-1}$, M $\approx$37~M$_{\sun}$; Martins, Schaerer \& Hillier~\cite{sl_martins05}), we estimate that of order 18 such stars would have been produced over 100~Myr given our CMD-derived SFR.  Of these, only a small fraction would be expected to be observed today owing to their short lifetimes. As an additional way to estimate the number of massive stars present, we employ a Monte Carlo approach to estimate the variance in massive star numbers given an overall low SFR. To that end, we assume a Salpeter IMF (Salpeter~\cite{sl_salpeter}) with lower and upper mass limits of 0.5 to 150~M$_{\sun}$, and draw from this distribution until 2.7$\times$10$^{5}$~M$_{\sun}$ total mass of stars are formed; this number corresponds to our recent SFR integrated over 10$^8$ yr. The process is repeated 100 times, allowing us to calculate that on average 1216$\pm$40 stars with masses appropriate to O stars (here assumed to be 18--100~M$_{\sun}$) are formed over this time period. At any given moment we would expect to see $\approx$120 O stars if the IMF was being fully sampled, assuming an average lifetime of 10~Myr.  The fact that we observe about an order of magnitude less confirms that SDIG is in a ``down'' state of star formation lasting at least the duration of a typical O star lifetime.  However, given the SFR determined over our youngest age bin (4--30~Myr) is higher than the lifetime average for the galaxy and the high gas content of SDIG, it seems more likely that this represents a combination of short-term SFR fluctuation and stochastic effects due to IMF sampling than to any physically meaningful suppression of star formation on longer timescales.

Because H$\alpha$ is known to be problematic in cases of very low SFR, the FUV emission provides a better estimator of the recent SFR, sampling stars down to a few solar masses, and ages of order 100~Myr (Lee et al.~\cite{sl_lee11}). We use the FUV asymptotic magnitude from Table~2 of Lee et al.~\cite{sl_lee11} in conjunction with the FUV-SFR relation from Lee et al.~(\cite{sl_lee09}; their $\S$3) to estimate a recent SFR of 9.6$\times$10$^{-4}$~M$_{\sun}$~yr$^{-1}$. For comparison, the SFR we derive from the CMD is 2.7$\pm$0.5~$\times$10$^{-3}$~M$_{\sun}$~yr$^{-1}$. We therefore find improved agreement between the recent SFR indicator and the CMD modelling, but a factor of 2--3 discrepancy remains, presumably due to the fact that a galaxy that is deficient in O stars will also be deficient in FUV flux.

\section{Summary and discussion}

  SDIG is a gas-rich, low-luminosity dIrr (C\^ot\'e et al.~\cite{sl_cote97}; Heisler et al.~{\cite{sl_heisler97}}). It is dominated by intermediate-age stars with an average age ${\rm \tau}=$6.4$^{+1.6}_{-1.4}$~Gyr, similar to the mean mass-weighted age of LG dIrrs (e.g., Orban et al.~\cite{sl_orban08}). The average SFR is similar to other Sculptor group dwarfs, listed in Weisz et al.~(\cite{sl_weisz11b}; their Table~2), with ESO540-G030 and ESO540-G032 having the lowest and highest average SFRs, respectively. The non-detection of $\ion {H}{ii}$ regions and low H$\alpha$-based SFR have been regarded as indicative of SDIG experiencing a currently quiescent phase (Miller \cite{sl_miller96}; Heisler et al.~\cite{sl_heisler97}; Skillman et al.~\cite{sl_skillman03}; Bouchard et al.~\cite{sl_bouchard09}). However, our SFH modelling shows elevated star formation for ages younger than 100~Myr, consistent with the star formation levels detected in the FUV emission. Our current data do not allow us to resolve star formation events lasting $\sim$10$^8$~Myr for ages older than $\sim$1~Gyr.  There are intriguing hints of a strong previous enhancement of SFR at ages of 600--1100~Myr, although the amplitude and timing of this event are model-dependent. Similar patterns of globally decreasing SFR during the past 1--2~Gyr have been observed in a number of LG galaxies with similar luminosity to SDIG, e.g. Leo~A (Cole et al.~\cite{sl_cole07}).

  SDIG is a member of the NGC\,7793 subgroup located at the far-side of the Sculptor group (Jerjen et al.~\cite{sl_jerjen98}; Karachentsev et al.~\cite{sl_karachentsev06}), and NGC\,7793 is identified as its main disturber, i.e. the neighbouring galaxy producing the maximum tidal action (Karachentsev et al.~\cite{sl_karachentsev04}). NGC\,7793 is a spiral galaxy and stellar population analyses of several halo and disk fields detect a metallicity gradient (Vlajic et al.~\cite{sl_vlajic11}), as well as indications for substantial stellar radial migration (Radburn-Smith et al.~\cite{sl_radburn12}). The SFH of a halo field in NGC\,7793 indicates that a burst of star formation has occurred within the last 20~Myr and with an amplitude three times that of the average SFR, most likely associated with the spiral arm feature contaminating the halo field (Dalcanton et al.~\cite{sl_dalcanton12}; Figs.~1, 2, and 9 for NGC\,7793). The deprojected distance between SDIG and NGC\,7793 is 450$^{+340}_{-270}$~kpc, estimated using the method in Karachentsev et al.~(\cite{sl_karachentsev04}), i.e. ${\rm R}^{2}={\rm D}^{2} + {\rm D}^{2}_{{\rm MD}} - 2 {\rm D} {\rm D}_{{\rm MD}}cos\Theta$, where ${\rm D}_{{\rm MD}}$ is the distance in Mpc of NGC\,7793 adopted from Vlajic et al.~(\cite{sl_vlajic11}), ${\rm D}$ is the distance in Mpc of SDIG, and $\Theta$ is their angular separation, calculated using NASA/IPAC Extragalactic Database (NED). The uncertainty in the deprojected distance takes into account the uncertainties in the distance of NGC\,7793 and SDIG, as well as their angular projected separation as the minimum physical distance between them. We note that many gas-rich irregular and transition dwarfs are similarly distant from the Milky Way, for example NGC\,6822 or Leo\,T (McConnachie 2012).  Given the mass of NGC\,7793 from Carignan \& Puche (\cite{sl_car90}) the orbital timescale for SDIG around NGC\,7793 would appear to be too long to be responsible for star formation activity on timescales shorter than a Hubble time, but other signs of disturbance are present.

  C\^ot\'e et al.~(\cite{sl_cote00}) discuss how the low value of SDIG's maximum rotational velocity implies that random motions may significantly contribute to the dynamical support and may thus lead to the puffing-up of an initially thin disk. To that end, the twisted {\ion {H}{i}} disk of SDIG (C\^ot\'e et al.~\cite{sl_cote00}) may indicate an interaction having occurred in this dwarf's past. NGC\,7793 has been mapped with Chandra (Pannuti et al.~\cite{sl_pannuti11}), and the results indicate an asymmetry in the detected X-ray sources. Pannuti et al.~(\cite{sl_pannuti11}) discuss an interaction between SDIG and NGC\,7793 as a plausible way to explain this asymmetry in the X-ray sources, although they do not regard it as strong a possibility. Radburn-Smith et al.~(\cite{sl_radburn12}) discuss that the detected stellar radial migration in NGC\,7793 may also be due to a past interaction event having occurred in NGC\,7793. While the deprojected distance of SDIG shows its current location and its actual orbit remains unknown (e.g. Bellazzini et al.~\cite{sl_bellazzini96}), the timescale of the star formation enhancements and the crossing timescale of SDIG around NGC\,7793 suggest that the elevated SFRs in SDIG are not due to gravitational interactions between them. The SF enhancements and disturbed {\ion {H}{i}} kinematics seen in SDIG may be due to the effect of gravitational interactions with a lower luminosity, gas-poor dwarf galaxy companion that has thus far remained elusive, as such tidal interactions between dwarf galaxies now start to become revealed (e.g., Martinez-Delgado et al.~\cite{sl_delgado12}; Rich et al.~\cite{sl_rich12}; Hunter et al.~\cite{sl_hunter98}). Tidal interactions between dwarf galaxies may also provide an important link to understanding the evolution and characteristics of dwarf galaxies in the isolated far-outskirts of the LG, as the case of the transition-type dwarf galaxy VV124 shows (e.g., Kirby et al. 2012).

  It is interesting in this respect that based on photometric arguments alone, Heisler et al.~(\cite{sl_heisler97}) conclude that it is not possible to rule out that SDIG-like systems may evolve to dSphs. It is thus intriguing that the off-center local sites of star formation observed in the young MS stars of SDIG, as well as other LG dwarfs (Dohm-Palmer et al.~\cite{sl_dohm-palmer97}; Gallagher et al.~\cite{sl_gallagher98}; Mateo \cite{sl_mateo98}; Cole et al.~\cite{sl_cole99}; McConnachie et al.~\cite{sl_mcconnachie06}), are observed in the majority of the transition-type dwarfs in the Sculptor group (Lianou et al.~\cite{sl_lianou12}; Jerjen \& Rejkuba \cite{sl_jerjen01}). Environmental effects on gas-rich dwarfs provide a possible avenue to their evolution into gas-poor systems (e.g., Mayer \cite{sl_mayer10}; and references therein), and such environmental effects seem to affect their recent star formation activity (e.g., Bouchard et al.~\cite{sl_bouchard09}; Thomas et al.~\cite{sl_thomas10}; Weisz et al.~\cite{sl_weisz11b}). However, the detection of a recent modest upturn in SFR, and hints of a similar episode $\approx$0.6-1~Gyr ago is further indication that small galaxies are able to sustain complex star formation histories without strong tidal interactions from larger neighbours.

\begin{acknowledgements} The authors would like to thank an anonymous referee for the thoughtful comments. We are grateful to Margrethe Wold and Claus Leitherer for useful discussions on Starburst 99. \\
  This research has made use of the following facilities: NASA/IPAC Extragalactic Database (NED) which is operated by the Jet Propulsion Laboratory, California Institute of Technology, under contract with the National Aeronautics and Space Administration; NASA's Astrophysics Data System Bibliographic Services; SAOImage DS9 developed by Smithsonian Astrophysical Observatory; Aladin.
\end{acknowledgements}

\end{document}